\documentstyle[12pt,epsfig]{article}
\newcommand{\ber}{\begin{eqnarray}}
\newcommand{\eer}{\end{eqnarray}}
\newcommand{\bea}{\begin{equation}}
\newcommand{\eea}{\end{equation}}
\newcommand{\del}{\partial}
\begin{document}
\title{\bf Quantum Simulation of non-Born-Oppenheimer dynamics in molecular systems by path integrals \\}
\author{\bf Sumita Datta $^{1,2}$\\
$^1$ Alliance School of Applied Mathematics, Alliance University,\\ Bengaluru 562 106, India\\
$^2$ Department of Physics, University of Texas at Arlington,\\Texas 76019, USA\\}
\maketitle
\begin{abstract}
A  numerical algorithm  based on the probabilistic path integral approach for solving Schr\"{o}dinger equation has been devised to treat molecular systems without Born-Oppenheimer
approximation in the  nonrelativistic limit at zero temperature as an alternative to conventioanl variational and perturbation methods. Using high quality variational trial functions and path integral method based on Generalized Feynman-Kac method, we have been able to
calculate the non-Born-Oppenheimer energy for hydrogen  molecule for the $X^{1}{{\Sigma}^1}_{g}$ state and hydrogen molecular ion. Combining 
these values and the value for ionization potential for atomic hydrogen, the dissociation energy and ionization potential for hydrogen molecules have been determined to be 36 113.672(3) $cm^{-1}$ and 124 446.066(10) $cm^{-1}$ respectively. Our  results favorably compare  with other theoretical and experimental results and thus show the promise of being a nonperturbative alternative for testing fundamental physical theories.
\end{abstract}
\newpage
\section{Introducton}
Sovling eigenvalue problems for molecular systems are in general complicated and people quite often make several simplications to handle
the difficulties associted with it, namely Born-Oppenheimer(BO) approximation or adiabatic  Born-Oppenheimer(ABO)[1,2,3] and rovibrational approach[4].
In the Born-Oppenheimer(BO) approximation in a chemical system, the coupling between the nuclear and electronic movements is neglected. The large mass disparity in the nuclei and electrons justifies the decoupling of their
different time scale motions and provides a very practical way to model a lot of chemical system adequetely.
In quantum mechanics, a solution of Schr\"{o}dinger equation in the  adiabatic approximation is defined to have a time scale separation in the fast and slow degrees of motion. Sometimes  the  BO approximation is also referred to as 'Adiabatic Born-Oppenheimer' approximation as the lighter electrons follow the motion of the heavier nulei adiabatically.
The ABO approximation breaks down
when two or more potential energy surfaces approach each other or cross and one must take resort to the coupled equations.
It is justifiable when energy gap between the ground and excited states is larger than the energy scale of nucleus. In metals, the  applicability of ABO is quaestionble as this energy gap turns out to be zero. For example, ABO fails in the case of Graphene[5].

With the typical BO approximation in a diatomic molecule the  non-relativistic ground state energy can be evaluated by solving the relevant
Schr\"{o}dinger equation neglecting the nuclear kinetic energy. One can choose some fixed value for the nuclear confugurations  and solve for the electronic wavefunction which depends parametrically on this fixed value of nuclear configuration which we will describe in the next section.
Since the nuclear configuration is considered  as a parameter and  not a quantum mechanical variable the nuclear motion is modelled only classically.
To get the full quantum dynamics of the diatomic molecule one needs to treat molecular systems as a whole, including electrons and atom nuclei on the same footing. Or in ther words one needs to consider
the motion of all the constituents of the molecule simultaneously assuming nuclei have finite masses and they move in the configuration space
as freely as the electrons do.
In view of the above reasons in this paper we have taken a relook at the sigma state of hydrogen molecule  and hydrogen ion molecule as a testbed so explore the fully non-Born-Oppenheimer(nBO) or non-adiabatic efects in diatomic molecular systems in general. As a matter of fact we have adopted a quantum Monte Carlo method based on Generalized Feynman-Kac(GFK) method[6-11] to calculate the energies  for the sigma state of hydrogen molecule and molecular hydrogen. Since GFK is a non-perterbative approach, it is easier to study motion of all the particles in the molecular system even in a fully quantum mechanical scenario. In the framework of GFK the non-BO study of hydrogen molecule now boils down to solving a four particle Schr\"{o}dinger equation. In the Stochastic scenario, the
all four particles execute twelve dimensional random walk scaled differently due to mass disparity of electrons and nuclei. In the case of a hydrogen molecular ion it turns out to be a nine dimensional walk of three particles(one electron and two nuclei). Adopting this idea  we prescribe the nBO model  for any aribitrary many body system with more than one electron and nucleus. Now to test the power of our theory we calculate the total energy of hydrogen molecular system(hydrogen molecule and hydrogen molecular ion) using path integral Monte Carlo technique use those to determine the ionization potential $E_p$ and dissociation energy $E_d$ of hydrogen molecule. At this point we need to review the theoretical and experimental endeavors and their outcomes so far.  Hydrogen molcule and hydrogen molecular ion are well explored topics in quantum mechanics. Their long history dates back to the first theoretical work of Heitler and London[12] followed by the work of James and Coolidge[13,14] and extension by  Kolos and Wolniewicz[15-17]. The controversy of [16,17,18] was apparently resolved by the experimental results of Hertzberg[19] and Stwalley[20]. Subsequent theoretical[21,22,23] and experimental endeavors[24,25,26] seemed to reduce the discrepency between the theory and experiment. Up to 2017 the best accepted theoretical value of the dissociation energy of hydrogen molecule in the non-BO basis is 36,118.0695(10)$cm^{-1}$[Piszczatowski et al [27],Pachuki et al [28] and Puchalski[29] and the best experimental value was 36,118.0696(4)$cm^
 {-1}$[Liu et al[30] and Altmann et al[31]. One of the most recent experimental works[32] has reported the  values for ionization potentials($E_p$) and dissociation energies($E_d$)for hydrogen molecule which are significantly lower than the accpted values so far. Most of the theoretical approaches[33-39] were based on variational principle and provided the sophistcated upper bound to the total energies, dissociation energy and ionization energy etc.  In this work using Non-BO basis functions as the trial functions[40] our path integral
approach yielded the nonrelativistic non-BO energies for both hydrogen molecule and molecular ion which are lower than the previously accepted values. These ground state energies for the hydrogen molecular system were used in calculating the ionization potentials($E_p$) and dissociation energies($E_d$) for hydrogen molecule and we get  new benchmarks for those.
 
The orgnization of the paper is as follows: In Section 1, we introduce the problem and describe the contents of the different sections of the 
paper. In Section 2.1, we discuss the general aspects of BO and nBO approaches to Quantum Mechanical problems. In Section 2.2 we discuss our path integral approach for calculating the eigen energies of hydrogen molecular system. In Section 3, we show how it can be generalized to any N particle molecular systems. In Section 4 we discuss our results and we conclude in Section 5. Fig 1 and 2 are schematic diagrams for the overall motion of the molecule in the nBO scenario. Fig 3 show how our numerical code works in the BO limit. 

\newpage
\section{Theory}
\subsection{General theoretical considerations in BO and nBO approach in connection with hydrogen molecular system}
In the BO approximation the nuclei are clamped at a fixed position and  
only electrons are moving in the configuration space. In the following Figs 1 and 2 the arrows beside the electrons as well as the nuclei signify that in the non-
BO approach electrons and nuclei all are moving in the configuration space and they are being treated in the same footing.
\begin{figure}[h!]
\includegraphics[width=6in,angle=-0]{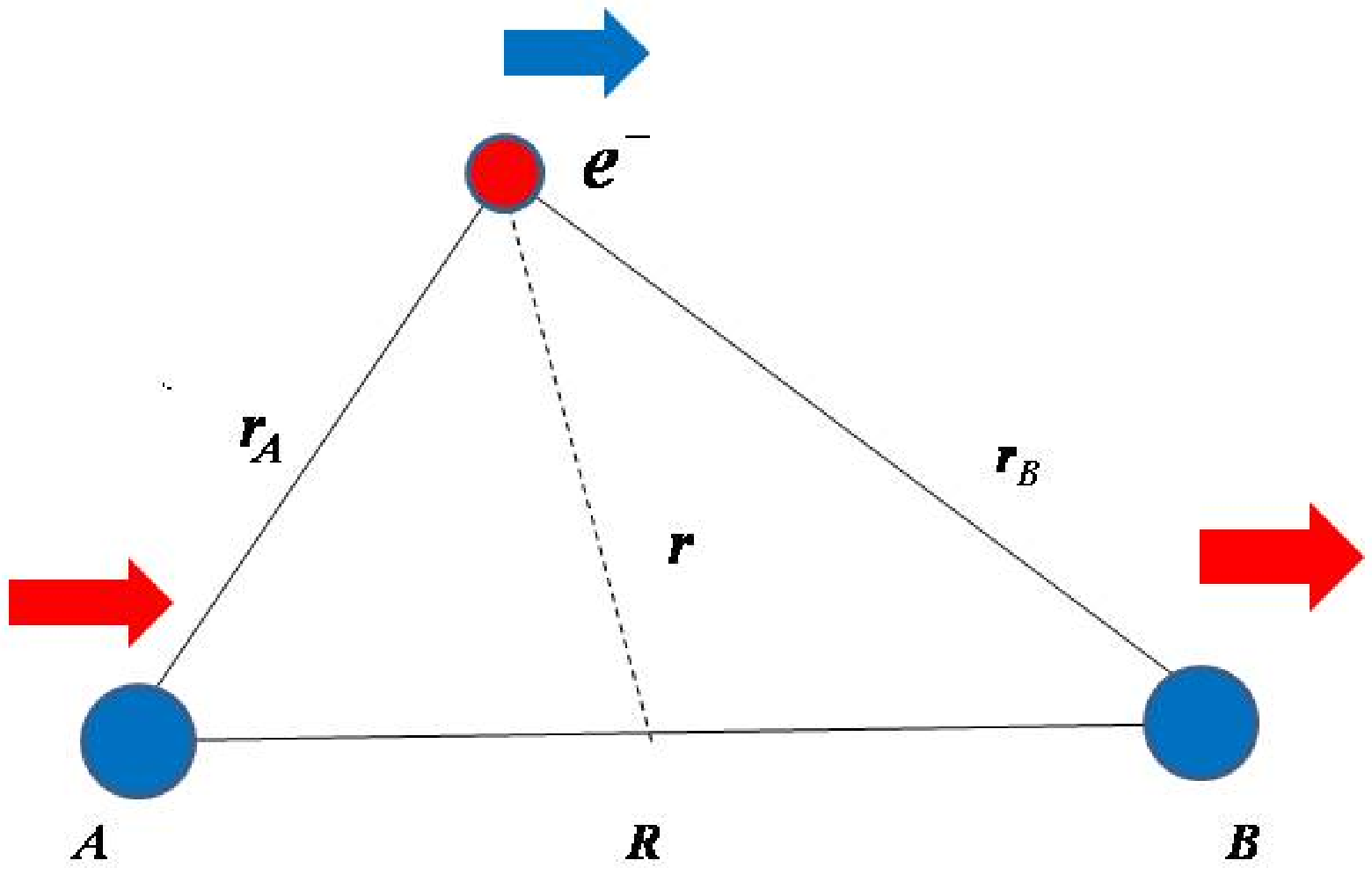}
\setlength\abovecaptionskip{-1.0\baselineskip}
\caption {A plot for the nBO dynamics of the hydrogen molecular ion} 
\end{figure}
\begin{figure}[h!]
\includegraphics[width=6in,angle=-0]{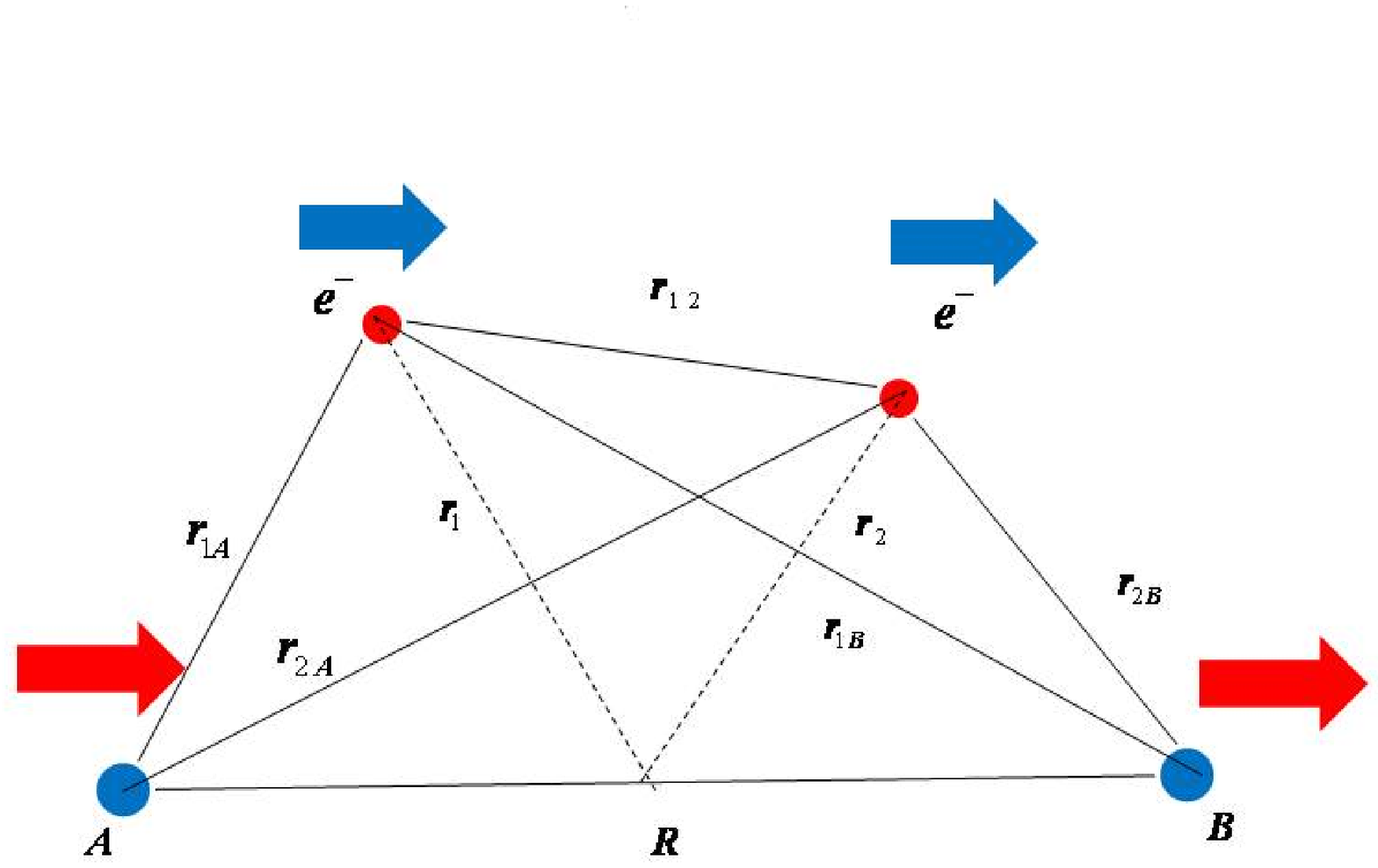}
\caption{A plot for the nBO dynamics of the hydrogen molecule}
\end{figure}
The exact time dependent Schr\"{o}dinger equation for hydrogen molecule can be represented by
$H\Psi(r_1,r_2,R_1,R_2)=E\Psi(r_1,r_2,R_1,R_2)$
with
\bea
H=\sum_{{\alpha}=A,B}-\frac{{{\nabla}_{\alpha}^2}}{2\frac{M}{m}}-\sum_{{\beta}=1,2}-\frac{{{\nabla}_{\beta}^2}}{2}-\frac{1}{r_{1A}}-\frac{1}{r_{1B}}
-\frac{1}{r_{2A}}-\frac{1}{r_{2B}}+\frac{1}{r_{12}}+\frac{1}{R}\\ \nonumber
\eea
where $r_{\beta},{\beta}=1,2,R_{\alpha},{\alpha}=A,B$ are the positions of the electrons and the nuclei respectively. As a matter of fact $r_{1A}=r_1-R_A,r_{1B}=r_1-R_B $ and so on $ r=r_1-r_2,R=R_A-R_B$.
 In defining the above Hamiltonian we have used atomic units and m and M denote the mass of electrons and nuclei respectively. Also $\frac{M}{m}=1836.152701$. The above Schr\"{o}dinger equation describes the quantum motion of the four constituents of hydrogen  molecule(two electrons and 2 nuclei).
For the hydrogem molecular ion the Schr\"{o}dinger equation can read as
\bea
H\Psi(r_A,r_B,R)=E\psi(r_A,r_B,R)
\eea
\bea
H=\sum_{\alpha=A,B}-\frac{{{\nabla}_{\alpha}^2}}{2\frac{M}{m}}-\frac{{{\nabla}^2}}{2}-\frac{1}{r_{A}}-\frac{1}{r_{B}}
+\frac{1}{R}\\ \nonumber
\eea
In the case of Born-Oppenheimer approximation, corresponding to the general Hamiltonian for the electron-nuclei system of hydrogen molecule 
$H(R,r)=T_R+H_e(R,r)$ there will be only one potential surface defined by $V(R)=<{\psi}_0|H_e(R,r)|{\psi}_0>$(as evident from Fig 3), whereas in the case of a non-BO calculation there will be multiple potential surfaces[41] corresponding to $V_{ii}(R)=<{\psi}_i|H_e(R,r)|{\psi}_i>$ and $V_{ij}(R)=<{\psi}_i|H_e(R,r)|{\psi}_j>$. 
\begin{figure}[h!]
\includegraphics[width=4in,angle=-90]{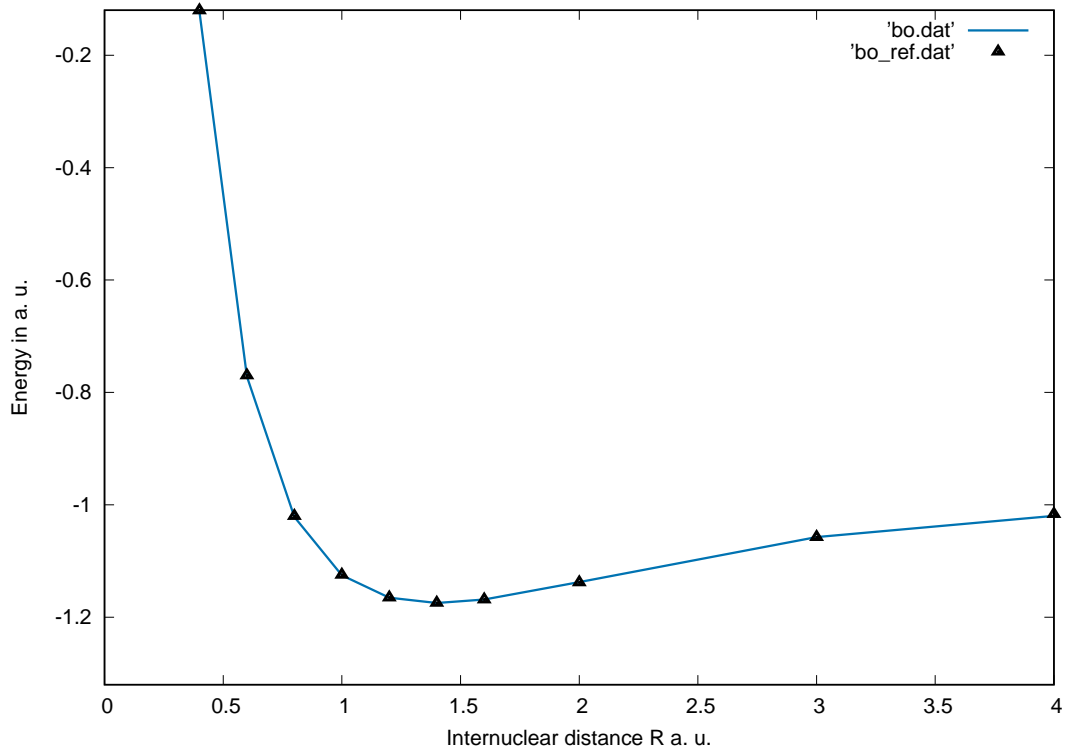}
\caption{A plot for the BO energy vs Internuclear distances}
\end{figure}
\clearpage
\subsection{ General Path Integral Theory for the Energy and other Properties for many body systems}
Let us consider the time-dependent Schroedinger
 equation for a system of N particles with Hamiltonian $H=-{{\nabla}^2}/2+V(x)$ as follows:
\ber
& & i\frac{\del {\psi}(x,t)}{\del t}=(-\frac{{\nabla}^2}{2}+V){\psi}(x,t) \\ \nonumber
& & \psi(x,0)=f(x)
\eer
where  $x\in R^d$.
The above initial value problem in imaginary time can be represented as 
\ber
& & \frac{\del {\psi}(x,t)}{\del t}=(\frac{{\nabla}^2}{2}-V){\psi}(x,t) \\ \nonumber
& & {\psi}(x,0)=f(x) 
\eer
The solution of Eq(5) for $f(x)=1$ can be written in the following Feynaman-Kac 
representation[42-43] and it provides a rigorous justification unlike ordinary Feynman path integration[44-47]. 
\bea
\psi(x,t)=E_{x}exp\{-\int_{0}^{t}V(X(s))ds\}
\eea
for $V\in K_{\nu}$, the Kato class of potential[48]
where $x(t)$ is  a Brownian motion trajectory and E is the average value of the exponetial term with respect to these trajectories. The lowest eigenvalue for a given symmetry can be obtained by large deviation principle
of Donsker and Varadhan[49],
\bea
{\lambda}_1=-\lim_{t\rightarrow \infty}\frac{1}{t}lnE_x[{e^{-\int_{0}^{t}V(X(s))ds})}]
\eea
The above
representation(Eq[6]) suffers from poor convergence rate as the underlying diffusion process-Brownian motion(Wiener Process[50] is non-recurrent. So it is necessary to use a representation which employs a diffusion which unlike Brownian motion, has a stationary distributions. To speed up the calculation we use generalized Feynman-Kac method in which the  Hamiltonian is rewritten as $H=H_0+V_p$ where $H_0=-{{\nabla}^2}/2+{\lambda}_T+{\nabla}{\psi}_T/2{\psi}_T$. Here ${\psi}_T$ is a  twice differentiable nonnegative reference function and $H{\psi}_T={\lambda}_T{\psi}_T$.
The expression for the energy can now be rewritten as 
\bea
{\lambda}_1={\lambda}_T-\lim_{t\rightarrow \infty}\frac{1}{t}lnE_x[{e^{-\int_{0}^{t}V(Y(s))ds})}]
\eea
where $Y(t)$ is the diffusion process which solves the stochastic differential equation and is known as Ornstein-Uhlenbeck process.
\bea 
dY(t)=\frac{\nabla {\Psi}_T(Y(t))}{{\Phi}_T(Y(t))}+dX(t) \\
\eea
$V(Y(s))$ is summed over all the time steps and $e^{-V(Y(s))}$ is summed over all the trajectories.
The presence of both drift and diffusion terms in the above expression enables the trajectory to be highly localized. As a result, the important regions of the potential are frequently sampled and Eq(4) converges rapidly.
The expectation value for the other properties can be evaluated as follows[6,51]:\\
\bea
{\langle Y|A|Y\rangle}=
\frac{\lim_{t\to\infty}\int dY(t)A(Y(t))e^{-\int[{V}_p(Y(s)]ds}}
{\int dY(t)e^{-\int[{V}_p(Y(s)]ds}}\,.
\label{former_13}
\eea
\section{Calculation BO and nBO energies and properties}
The nonrelativistic Hamiltonian for hydrogen molecule can be written as a sum of five terms:
\bea
=-\frac{{\hbar}^2}{2M} \sum_{\alpha}{\nabla}_{\alpha}^2-\frac{{\hbar}^2}{2m} \sum_{\beta}{\nabla}_{\beta}^2-\sum_{{\alpha},{\beta}}
\frac{e^2}{4\pi{\epsilon}_{0}r_{{\alpha}{\beta}}}+\sum_{A>B}\frac{e^2}{4\pi{\epsilon}_{0}R_{AB}}+\sum_{{\beta}>{\gamma}}\frac{e^2}{4\pi{\epsilon}_{0}r_{{\beta}{\gamma}}}
\eea 
In atomic units the above expression is just
\bea
H=-\frac{1}{2M}\sum_{\alpha}{\nabla}_{\alpha}^2-\frac{1}{2}\sum_{\beta}{\nabla}_{\beta}^2-\sum_{{\alpha},{\beta}}\frac{1}{r_{{\alpha}{\beta}}}+\sum_{A>B}\frac{1}{R_{AB}}+\sum_{{\beta}>{\gamma}}\frac{1}{r_{{\beta}{\gamma}}}
\eea
For Born-Oppenheimer calculations the kinetic energy of nuclei is neglected due to its small contribution  towards the total energy as it gets  divided by 'M', the mass of nuclei. So within BO approximation,the above equation reads as follows:
\bea
H=-\frac{1}{2}\sum_{\beta}{\nabla}_{\beta}^2-\sum_{{\alpha},{\beta}}\frac{1}{r_{{\alpha}{\beta}}}+\sum_{A>B}\frac{1}{R_{AB}}+\sum_{{\beta}>{\gamma}}\frac{1}{r_{{\beta}{\gamma}}}
\eea
Here ${R_A} $ and ${R_B} $ are not variables hence $R_{AB}$ is treated as a parameter.\\
In general the Hamiltonian for all the moving electrons and the nuclei has the form
\ber
 H= -\frac{1}{2m}{\nabla}^2(r_1,r_2)-\frac{1}{2M}{\nabla}^2(R_A,R_B) \nonumber \\
+\frac{1}{|r_1-r_2|}+\frac{1}{|R_A-R_B|}  \nonumber \\
-\frac{1}{|r_1-R_A|}-\frac{1}{|r_1-R_B|}-\frac{1}{|r_2-R_A|}-\frac{1}{|r_2-R_B|} \nonumber \\
 =  -\frac{1}{2}{\nabla}^2+V(r_1,r_2,R_A,R_B)
\eer
where $r_{\beta},{\beta}=1,2,R_{\alpha},{\alpha}=A,B$ are the positions of the electrons and the 'quantum nuclei'(moving nuclei) respectively(Fig 2). $|.|$ is the 3 dimensional Euclidean distance and $V$ is the Coulombic interaction.
For non-Born-Oppenheimer calculations the hydrogen molecules can be  treated as general physical systems with  inequivalent masses(electrons and nuclei)  and can be represented with the above Hamiltonian.
For any physical system with N ineqivalent masses the above can be generalized as follows:
The  Schr\"{o}dinger equation for the above system can be written as
\ber
& &[-\sum_{i}\frac{{\hbar}^2}{2{m_i}}{\nabla}^2(\vec{r}_i)-\sum_{j}\frac{{\hbar}^2}{2{M_j}}{\nabla}^2(\vec{R}_j) \nonumber \\
& &+\sum_{ik}\frac{1}{\sqrt{(r_i-r_k)^2}}+\sum_{jl}\frac{1}{\sqrt{(R_j-R_l)^2}}
-\sum_{ij}\frac{1}{\sqrt{(R_j-r_i)^2}}]{\psi}({\vec{r}}_i,{\vec{R}}_j)
=\mu {\psi}({\vec{r}}_i,{\vec{R}}_j) \nonumber \\
\eer
i,j refer to number of elctrons and number of nuclei respectively.
Now using
$\vec{r_i}=s_i\vec{r_i}^{\prime}$
$\vec{R_j}=s_j\vec{R_j}^{\prime}$,
${\nabla}^2(\vec{r}_i)={\nabla}^2(s_i\vec{r_i}^{\prime})
=\frac{1}{{s_i}^2}{\nabla}^2(\vec{r_i}^{\prime})$ \\
${\nabla}^2(\vec{R}_j)={\nabla}^2(s_j\vec{R_j}^{\prime})
=\frac{1}{{s_j}^2}{\nabla}^2(\vec{R_j}^{\prime})$
and putting
$m_i{s_i}^2=M_j{s_j}^2$, multiplying throughout by $\frac{m_i{s_i}^2}{\hbar^2}$ the Schr\"{o}dinger equation in the dimensionless form reads as
(in $m_i=\hbar=1$ units)
\ber
& &[-\sum_{i}\frac{{\nabla}^2(\vec{r_i}^{\prime})}{2}-\sum_{j}\frac{{\nabla}^2(\vec{R_j}^{\prime})}{2} \nonumber \\
& & +\sum_{ik}\frac{1}{\sqrt{(\vec{r_i}^{\prime}-\vec{r_k}^{\prime})^2}}+\sum_{jl}\frac{1}{\sqrt{(\vec{R_j}^{\prime}-\vec{R_l}^{\prime})^2}}-\sum_{ij}\frac{1}{\sqrt{(\frac{m_i}{M_j}\vec{R_j}^{\prime}-\vec{r_i}^{\prime})^2}}]{\psi}({\vec{r}^{\prime}}_i,{\vec{R}^{\prime}}_j)
=\mu {\psi}({\vec{r}^{\prime}}_i,{\vec{R}^{\prime}}_j) \nonumber \\
\eer
In general let $m_i{s_i}^2=M_j{s_j}^2$ and set $s_i=\sqrt{\frac{m_i}{M_j}}s_j$
No matter the number of distinct masses, the scale for the corresponding random walker will always be the square root of the ratio of their masses. Using this result, the Generalized Feynman-Kac path integral can be used for non adiabatic treatment of the molecules.
Now for hydrogen molecule we need to simulate the random ealk associated with the following Hamiltonian 
\ber
H =-\frac{1}{2}{\nabla}^2(r_1,r_2)-\frac{1}{2}{\nabla}^2(R_A,R_B) \nonumber \\ 
-\frac{1}{\sqrt{(\sqrt{\frac{m}{M}}X_A-x_1)^2 
  +(\sqrt{\frac{m}{M}}Y_A-y_1)^2  
+(\sqrt{\frac{m}{M}}Z_A-z_1)^2}} \\ \nonumber 
-\frac{1}{\sqrt{(\sqrt{\frac{m}{M}}X_B-x_1)^2
  +(\sqrt{\frac{m}{M}}Y_B-y_1)^2
 +(\sqrt{\frac{m}{M}}Z_B-z_1)^2}} \\  \nonumber 
-\frac{1}{\sqrt{(\sqrt{\frac{m}{M}}X_A-x_2)^2
  +(\sqrt{\frac{m}{M}}Y_A-y_2)^2
 +(\sqrt{\frac{m}{M}}Z_A-z_2)^2}} \\ \nonumber 
 -\frac{1}{\sqrt{(\sqrt{\frac{m}{M}}X_B-x_2)^2
  +(\sqrt{\frac{m}{M}}Y_A-y_2)^2
 +(\sqrt{\frac{m}{M}}Z_A-z_2)^2}} \\ \nonumber 
+\frac{1}{\sqrt{(X_A-X_B)^2+(Y_A-Y_B)^2+(Z_A-Z_B)^2}} \\ \nonumber 
 +\frac{1}{\sqrt{(x_1-x_2)^2+(y_1-y_2)^2+(z_1-z_2)^2}} 
\eer
\section{Results and discusssions}
Now by setting $x=\{x_k\},k=1,2,....,12,{x_k} {\bf{{\in}R^{12}}}$
we can write $V(x)=V(r_1,r_2,R_A,R_B)$
To calculate energies, we use Eq(8) of Sec 2.2 whereas the other properties are calculated using Eq(10) with $V$ as defined  in Eq(16). In our
program the stepsize is fixed and the direction of the path is chosen randomly.
In Eq(9),the first and second terms represent the drift and diffusion respectively. 
For each system a number of paths are generated with a specific path length. These are then summed to produce an average value and a statistical error. In order to examine the behavior of our energies and other properties as a function of path length, we compute several different path lengths - from 8 to 48 units of time. To implement Eq(8) numerically we replace 12 dimensional Brownian motion with 12 dimensional Ornstein-Uhlenbeck process and simulate them by 12 independent, properly scaled one dimensional drifted random walk. For details plesae see Appendix A . 
The Hamiltonian in Eq(17) is represented as twelve dimensional random walk where six dimensions involve stepsizes $\frac{m}{M}$ of the size of other six as shown above.
We calculate $E(H_2)$ and $E(H_2^+)$ using the formula given  Eqn(8). Using these values and the value of ionization potential of hydrogen one can calculate the  Ionization Potential and Dissociation Energy according to the following expressions[52].\\
{\bf Ionization Potential:}\\
\bea 
E_P=E(H_2^+)-E(H_2)\\
\eea
{\bf Dissociation Energy:}\\
\bea
E_d=-\lim_{R\rightarrow \infty}E(H_2)-E(H_2)\\ \nonumber
=-2E(H)-E(H_2)
\eea
For the BO and NBO calculations for the hydrogen molecule we use the trial function of the following form:
\bea
{{\psi}_T}=(1+P_{12})(1+P_{AB})exp({\sum}_{k=0}a_k{q_{1A}}^{u}{q_{1B}}^{v}{q_{2A}}^w{q_{2B}}^n{q_{12}}^g{q_{AB}}^h-\chi r_{1A}-\delta r_{2B})
\eea
Here $P_{12}$ is the operator that interchanges the two electrons, $P_{AB}$ is the operator that interchages the two nuclei, and
$q_x={r_x}/(1+c{r_x})$ is a coordinate transformation that allows terms in the exponent to go smoothly to the separated atom limit.
The exponents u, v, w, n, g,and h are integers(0,1,........) and all possible terms adding up to $N=u+v+w+n+g+h$ are selected.
In Table 2,  we show the variation of BO energies with internuclear distances and its agreement with the best nonrealtivistic estimates for this system. We show this agreement for BO case just to establish that the same code works when the nuclei are frozen as well. In Table 3, we show extrapolated values for the BO energies for hydrogen molecule and hydrogen molecular ion. In Table 4, we run simulations for ${H_2}^{+}$ for different time 8-48. In Table 5, we show the comparison of extrapolated energy from the data in Table 4 with the best non relativistic variational calculatios. Table 6 contains the simlations for nbo dynamics of $H_2$ molecule at different time (8-48). In Table 7 we show the extrapolated value of the nbo energy of hydrogen moecule and its agreement with other theoretical values. As can be seen from Eq(8) the most accurate estimate of the energy is obtained when we extrapolate our results to infinite time. We do this by performing a least square fit. We have verified that the 5000 path lengths selected with runtime selected from 8 to 48 are more than enough to perform an accurate fit. Oher tests have confirmed that a stepsize of 1/30 has little influence on the value of extrapolated energy. The final value obtained for the nBO energy for the hydrogen molecule,-1.164 546(3), is in excellent agreement with the best non relativistic value for this system[36,58]. It is, better than the value obtained from the Variational Monte Carlo calculation. 
This agreement can, however, be attributed to the quality of the original trial wavefunction. Other tests have confirmed that the stepsize used has little influence on these values. Unlike the energy there is no need to extrapolate any of the properties to infinite time
 In Table 8, we have compared our Ionization potential $E_p$ and dissociation energy $E_d$ with other theoretical and experimental results.\\
For the BO and NBO calculations for the hydrogen molecular ion the following trial function function is used:\\
\bea
{\phi}_T=exp(-\sigma r^2)\\
\eea
where $\sigma$ is a variational parameter.
Using the calculated values for $E(H_2^{+})$(The extrapolated nBO energy for hydrogen molecular ion from Table 5), $E(H_2)$ (The extrapolated nBO energy for hydrogen molecule Table 7)and Eq(18), we calculate the value for the ionization potential $E_p(H_2)=0.567 018 a. u.=124 446.0663 cm^{-1}$. Using ionization energy for atomic hydrogen $E_p(H)$, and $E_(H_2)$ and Eq(19), we calculate the dissociation energy $E_d(H_2)=0.164 546 a.u.=36 113.672 cm^{-1}$. Now adding the value of previously calculated relativistic corrections[53] for sigma state to our nBO energy we determine the $E_d(H_2)$ to be equal to 36,116.672(10).\\   
{\bf Acknowledgements:}
The author would like to thank  Alliance University for providing partial support for carrying the research work and The University of Texas
at Arlington,USA where the idea behind the research work was conceived. 
\section{Conclusions and Outlook}
In this paper using a probabilistic approach a solution to time dependent Schr\"{o}dinger equation has been constructed in the form of a path integral. By simulating an approximation to Ornstein-Uhenlenbech  process through trial function(drift term) and a toss of a unbiased coin(diffusion term), we have determined the ionization potential and dissociation energy of hydrogen molecule with a high accuracy. It looks very promising to observe that if we guide our random walk using a non-BO variational trial function and perform the numerical simulation for the path integarl solution for the molecular system, we already improve the variational energy for the hydrogen molecule. For implementing the simulation of our path integral solution we make only onefold approximation as the probalistic representation to Schr\"{o}dinger equation  can be written in a closed form.  To be precise to calculate energy we approximate an exact solution (i. e. the GFK representation of it) to the Schr\"{o}dinger equation, whereas most of the other numerical procedures approximate a solution to an approximate Schr\"{o}dinger equation. Also since the path integral solution is based on fully quantum mechanical approach it can improve the variational energy to a graet extent provided our trial function has the right symmetry of the Hamiltonian. At this point our dissociation energy a little less than established theoretical and experimental values. We believe we need to add a non-BO relativistic correction to our present nBO non-relativistic energy to have a better agreement with the experimental values. We see a better agreement if we add our BO relativistic corrections[53] as a rough  estimate of non-BO relativistic corrections. Also we need to improve the quality of the trial funtions particularly in the case of hydrogen molecular ion. 
The procedre can be applied to more complex systems for which energies are known up to a few significat figures from variational calculations and accuracy can be increased to include more significant figures. Our benchmarks for the dissociation energy and ionization potential can be a useful input for the other work for evaluating non-BO relativistic corrections for the molecular systems. We observe that adopting this Monte Carlo method and taking advantage of modern computer technology solving eigenvalue calculations in Quantum mechanics can be simplified to a great deal and hope it will inspire other people to carry out research along the same line.   

\newpage
\begin{table}[h!]
\begin{center}
\caption{\bf Notation Table}
\begin{tabular}{ccc}
Notation/Phrase & Meaning \\
$r_{\beta}$      &   position vector of electrons inside the atom \\
$R_{\alpha}$    &   position vector of nuclei \\
$R$            & distance between two nuclei \\  
$r$   &   distance between two electrons\\
$X(t)$     &   Brownian motion with a non-ergodic probabilistic measure or\\
           &    Wiener Measure\\
$Y(t)$     &   A stochastic process with an ergodic or stationary measure\\
$ \psi_T$  &   The  trial function corresponding to mathematical ground state\\
$\phi_{T}$ &  The trial function for the molecular ion\\
$\hat{T}$  & Kinetic energy operator\\
$\hat{V}$  & Potential Energy operator\\
$2<\hat{T}>=n<\hat{V}$ & Virial theorem\\
$E_p(H)$  & ionization potential of atomic hydrogen $=0.5 a. u.$ \\
$E(H_2)$  & Total energy for the hydrogen molecule\\
$E({H_2}^{+})$ & Total energy for the  hydrogen molecular ion \\
$E_d(H_2)$    & dissociation energy for the hydrogen molecule \\
$E_p(H_2)$    & ionization potential for the hydrogen molecule \\
\end{tabular}
\end{center}
\end{table}
\newpage
\begin{table}[h!]
\begin{center}
\caption {\bf Born-Oppenheimer Energy  of Hydrgen molecule for the ground state for different internuclear distances. The number in the parentheses is the statiscal error.}
\begin{tabular}{cccccc}
R & E &     Refs.\\
0.4 &-0.122 348(5)& This work\\
    &-0.120 230   & Sims et al[54]\\
    &-0.120 228 2(9)& Alexander et al[55]\\
0.6 &-0.771 535(1)& This Work  \\
    &-0.769 635   & Sims et al[54]\\
    &-0.769 635 1 (4) & Alexander et al[55] \\
0.8 &-1.021 424(6)& This Work  \\
    &-1.020 056   & Sims et al[54] \\
    &-1.020 056 1(4) & Alexander et al[55]\\ 
1.0 &-1.125 406   & This work  \\
    &-1.124 539   & Sims et al[54] \\
    &-1.124 539 2(4) & Aleaxnder et al[55]\\ 
1.2 &-1.165 377(1)& This work   \\
    &-1.164 935   & Sims et al[54]  \\
    &-1.164 934 8(5) & Aleaxnder at al[55]\\
1.4 &-1.174 564(2)& This work \\
    &-1.174 475   & Sims at al[54] \\
    &-1.174 475 4(6)& Aleaxnder et al[55]\\ 
    &-1.174 475 1(5) & Datta et al[56] \\
   & -1.174 447 477 & Kolos et al[17] \\
   & -1.174 475 686 & Kolos[57] \\ 
1.6 &-1.168 371(1)    & This work  \\
    &-1.168 583    & Sims et al[54] \\
    &-1.168 583 4(5) & Alexander et al[55]\\
2.0 &-1.137 488(2)    & This work  \\
    &-1.138 132    & Sims et al[54]\\
    &-1.138 132 0(4) & Aleaxnder et al[55]\\ 
3.0&-1.057 351(1)   & This work  \\
   &-1.057 326   & Sims et al[54] \\
   &-1.057 324 9(3)& Alexander et al[55]\\ 
4.0&-1.019 750(4)    & This work \\
   &-1.016 390   & Sims et al[54] \\
   &-1.016 389 2(2)& Alexander et al[55]\\ 
\end{tabular}
\end{center}
\end{table}
\begin{table}[h!]
\begin{center}
\caption {\bf Total nonrelativistic Born-Oppenheimer Energy of Hydrgen molecule ion  and hydrogen molecule for the ground state at equilibrium distances. The number in the parentheses is the statiscal error.}
\begin{tabular}{cccccc}
molecule&R & E \\
$H_2^+$ & 2.0  &-0.609 148(1) & This work\\
        & 2.01 &-0.602 1      & Swarwono et al[59] \\
$H_2 $  & 1.4  & -1.174 564(2)& This work \\
\end{tabular}
\end{center}
\end{table}
\begin{table}[h!]
\begin{center}
\caption{\bf Total non-relativistic non Born-Oppenheimer energy in (a.u.) of hydrogen molecular ion for the ground state with Generalized Feynman-Kac(GFK) path integral method at different  time with a stepsize of $1/30$ and 5000 paths.}
\begin{tabular}{ccccc}
Time & $E_{mol}(GFK)$ &   $<T>$    & $<V>$        & Virial ratio \\
8    &-0.554700(1)      & 0.59921  &-0.92368 &1.541   \\
16   &-0.553210(4)    & 0.78576  & -1.4218 & 1.809   \\
24   &-0.458169(2)    & 0.3784   &-0.6724  & 1.776   \\
32   &-0.431555(6)    & 0.7140   &-1.2891  &  1.805  \\
40   &-0.420000(4)    & 0.616   &-1.036   & 1.681  \\
48  & -0.376000(6)    & 0.417   &-0.793 & 1.9 \\
\end{tabular}
\end{center}
\end{table}
\newpage
\begin{table}[h!]
\begin{center}
\caption {\bf Total Nonrelativistic non-Born-Oppenheimer enegies $E_{ion}$ of Hydrgen molecule ion in the ground state. The number in the parentheses is the statistical error.}
\begin{tabular}{cccccc}
Work& Method & $E_{ion}$ (a.u.) \\
Yuh et al& free iterative complement method(variational)& -0.597 139 \\
Jeziorski et al& rel variational & -0.597 144 \\
present work & GFK & -0.597 528(2) \\
\end{tabular}
\end{center}
\end{table}
\clearpage
\newpage
\begin{table}[h!]
\begin{center}
\caption{\bf Total non-relativistic non Born-Oppenheimer energy in (a.u.) of hydrogen molecule for the ground state with Generalized Feynman-Kac(GFK) path integral method at different  time with a stepsize of $1/30$ and 10000 paths.}
\begin{tabular}{ccccc}
Time & $E_{mol}(GFK)$ &   $<T>$    & $<V>$        & Virial ratio \\
8    & -1.164376(5)  & 1.17594 & -2.159756  & 1.836  \\
16   & -1.159098(4)  & 1.136411 & -2.23381   & 1.97  \\
24   & -1.163043(1)  & 1.266482 & -2.275504  & 1.8  \\
32   & -1.165153(2)  & 0.9991487 & -1.90263  & 1.904 \\
40   & -1.164310(5)  & 1.235915  & -2.584469 & 2.09 \\
48  &  -1.15559(2)  & 1.120355  & -2.27630 & 2.03  \\
\end{tabular}
\end{center}
\end{table}
\qquad
\begin{table}[h!]
\begin{center}
\caption {\bf Total Nonrelativistic non-Born-Oppenheimer enegies$E_{mol}$ of Hydrgen molecule  in the ground state . The number in the parentheses is the statiscal error.}
\begin{tabular}{cccccc}
Work& Method & $E_{mol}$ (a.u.) \\
Tubman et al[58]&FN DMC-full &-1.164 01(5) \\
Bubin et al[36] & non-BO-Var & -1.164 0250 \\
Alexander et al[40] &non-BO-Var & -1.164 02491(8) \\
present work & GFK & -1.164 546(3) \\
\end{tabular}
\end{center}
\end{table}
\newpage
\newpage
\begin{table}[h!]
\begin{center}
\caption {\bf ionization energy $E_p$,dissociation energy $E_d$ of hydrogen molecule in ${cm}^{-1})$ for the ground state. The number in the parentheses is the statiscal error.}
\begin{tabular}{lllll}
Work & method&  $E_p$ & $E_d$ \\
Herzberg et al[18] & Expt     &                      & 36 113.6$\pm 0.3$\\
Herzberg[19]      & Expt      &                      & 36 116.3$<D_0<36.118.3$ \\ 
Wolneiwicz[57]    & Theo(BO)  & 124 417.491          &                         \\  
Zhang et al[26]   & Expt      &                      & 36 118.062(10) \\
Liu et al[30]     & Hybrid    &                      &                 \\ 
                  & Expt      &                      &                 \\
                  &-Theo      & 124 417.491 13(37)   & 36 118.069 62(37) \\
Altmann et al[31] & Expt      &                      & 36 118 06945(31) \\
Piszczatowski &               &                      &                  \\ 
et al[27]         & Var       &                      &                  \\ 
                  &(Theo)     &                      & 36 118.0695(10)\\
Stwalley[20]      & Expt      &                      & 36 118.6\\
Pachuki et al[28] & Var       &                      &           \\ 
             &(Theo)          &                      & 36 118.797 746 3(2)  \\
Puchalskii et al[29] & (Theo) &                      & 36 118.067 8(6) \\ 
Cheng et al[32]   & Expt      & 124 357.238062(25)   & 35 999.582 894 (25)\\
Wang et al[38]   & Var(Theo)  &                      & 36 118.069 71(33) \\ 
Present work  & GFK(Theo)(non rel)& 124 446.066 (10) & 36 113.672(3)\\
Present work  & GFK with rel correction  &           &              \\  
              &(energy nBO+rel corr BO)  &           & 36 116.072(10)\\ 
\end{tabular}
\end{center}
\end{table}
\clearpage
\newpage
\appendix
\setcounter{equation}{0}
\renewcommand{\theequation}{A\arabic{equation}}
\section{Details of Numerical Calculations}
The formalism described in section 2 can include any generalized potential \cite{60} and valid for any arbitrary
dimension d (d=3N). To implement Eq(3) numerically, the 3N dimensional Brownian motion can be replaced by
properly scaled one dimensional random walks as follows \cite{9,43,61}:
\ber
W(l)\equiv W(t,n,l)
& = & {w_1}^1(t,n,l),{w_2}^1(t,n,l),{w_3}^1(t,n,l)....\\ \nonumber
&   &                  .......{w_1}^N(t,n,l){w_2}^N(t,n,l){w_3}^N(t,n,l)
\eer
where
\bea
{w_j}^i(t,n,l)=\sum^l_{k=1}\frac{{\epsilon}^i_{jk}}{\sqrt n}
\eea
with ${w_j}^i(0,n,l)=0$
for $i=1,2,....,N$;$j=1,2,3$ and $l=1,2,.....,nt$. Here $\epsilon $ denotes the binomially distributed random variables which are
chosen independently and randomly with probability P for all i,j,k such that
$P({\epsilon}^i_{jk}=1)$=$P({\epsilon}^i_{jk}=-1)$=$\frac{1}{2}$. It is known
by an invariance principle \cite{62} that for every $\nu$ and W(l)
defined in Eq.(A1) and Eq(A2)
\ber
\lim_{n\to\infty}P(\frac{1}{n}\sum^{nt}_{l=1}V(W(l)))\leq \nu \\ \nonumber
 =  P( \int\limits^t_0 V( X(s))ds)\leq\nu
 \,\,.
\eer
Consequently for large n,
\ber
P[ \exp(- \int\limits^t_0 V(X(s))ds)\leq\nu ] \\ \nonumber
 \approx  P [\exp(-\frac{1}{n}\sum^{nt}_{l=1}V(W(l)))\leq \nu]
\eer
Finally, by generating $N_{rep}$ independent realization $Z_1$,$Z_2$,....$Z_{N_{rep}}$ of
\bea
Z_m=\exp(-(-\frac{1}{n}\sum^{nt}_{l=1}V(W(l)))
\eea
and using the law of large numbers,with regard to Eq(A3), we conclude that
\bea
(Z_1+Z_2+...Z_{N_{rep}})/N_{rep}=Z(t)
\eea
is an approximation to Eq.(6)
Here $W^m(l), m=1,2,N_{rep}$ denotes the $m^{th}$ realization of W(l) out of
$N_{rep}$ independently run simulations. In the limit of large t and $N_{rep}$
this approximation approaches an equality, and forms the basis of a
computational scheme for the lowest energy of a many particle system with
a prescribed symmetry.


\newpage
\begin{thebibliography}{99}
\bibitem{1}M. Born and R. Oppenheimer, Ann. Phys.,{\bf 84},457(1927)
\bibitem{2}M. Born, Festschrift Gott. Nachr. Math. Phys.,{\bf KI},1(1951)
\bibitem{3}M. Born and K. Huang, The dynamical Theory of Crystal Lattices, Oxford University Press, London(1954)
\bibitem{4}S. A. Alexander and R. L. Coldwell, J Chem Phys,{\bf 129},345(2007)
\bibitem{5}S. Pisana,M. Lazzeri, C. Casiraghi, K. S. Novoselov, A. K. Geim, A. C. Ferrrari and F. Mauri, Nature Material{\bf 6}, 198–201 (2007). https://doi.org/10.1038/nmat1846
\bibitem{6}M.Cafferel and P. Claverie, J. Chem Phys. {\bf 88 }, 1088 (1988);{\bf 88}, 1100 (1988).
\bibitem{7}F. Soto-Eguibar and P Claverie,in Stochastic Processes Applied to Physics and other A Rueda(World Scientific, Singapore,1983).
\bibitem{8} A. Korzeniowski, J Comp and App Math {\bf 66}, 333 (1996)
\bibitem{9} S. Datta, J. L Fry, N. G. Fazleev, S. A. Alexander and R. L. Coldwell, Phys Rev A {\bf 61} R030502 (2000); 
S. Datta, Ph. D dissertation, The University of Texas at Arlington (1996).
\bibitem{10}S. Datta, Int. J. Mod. Phys.B,{\bf 37},2350024(2023)
\bibitem{11}S. Datta, V. Dunjko, M. Olshanii, Physics, {\bf 4}, 12, 2022
\bibitem{12}W.Heitler, F. London, Z. Phys.,{\bf 44}455,1927
\bibitem{13}H. M. James and A. S. Coolidge,{\bf 1},825,1933
\bibitem{14}H. M. James and A. S.Coolidge,{\bf 3},129,1935
\bibitem{15}W.Kolos and L. Wolniewicz, J. Chem. Phys.,{\bf 41},3663(1964)
\bibitem{16}W.Kolos and L. Wolniewicz, Phys. Rev. Lett.,{\bf 20},243(1968)
\bibitem{17}W.Kolos and L. Wolniewicz, J. Chem. Phys.,{\bf 49},404(1968)
\bibitem{18}G Herzberg and A. Monfils, J Mol. Spectr.{\bf 5},482(1960) 
\bibitem{19}G. Herzberg, J. Mol. Spectr,{\bf 33},147(1970)
\bibitem{20}W. C. Stwalley, Chem. Phys. Lett.,{\bf 6},241(1970)
\bibitem{21}W. Kolos, K. Szalewicz and H. J. Monkhorst, J. Chem.Phys., {\bf 84},3278(1986)
\bibitem{22}W. Kolos and J. Rychlewski,J. Chem. Phys.,{\bf 98},3960(1993)
\bibitem{23}L.J. Wolniewicz, J. Chem. Phys., {\bf 103},1792(1995)
\bibitem{24}A. Balakrishnan, V. Smith and B. P. Stoicheff, Phys. Rev. Lett.,{\bf 68},2149(1992)
\bibitem{25}E. E. Eyler and N. Melikechi,Phys.Rev. A,{\bf 48},R18(1993)
\bibitem{26}Y. P. Zhang, C. H Zhang, J. T. Kim, J Stanojevic and the E. E. Eyler, Phys. Rev. Lett. {\bf 92}, 203003(2004)
\bibitem{27}K. Piszczatowski, G.Lach, B.Jeziorski, M. Przybytek, J Komasa and K. Pachuki, J. Chem. Theo. and computation,{\bf 5},3039(2009)
\bibitem{28}K. Pachuki and J. Komasa, J. Chem. Phys.,{\bf 144}, 164306(2016)
\bibitem{29}M. Puchalski and J. Komasa, Phys. Rev. A,{\bf 95}, 052506(2017)
\bibitem{30}J. Liu, E. J. Salumbides, U. Hollenstein, J. C. J. Koelemeij, K. S. E. Eikema, W. Ubachs and F. Merkt, J. Chem Phys.,{\bf 130}174306,(2009)
\bibitem{31}R. K. Altmann, L.S. Dressen, E. J. Salumbides, W. Ubachs, K. S. E. Eikema, Phys. Rev. Lett,{\bf 120}, 043204(2018)204(2018)
\bibitem{32}C. Cheng, J. Hussels, M. Niu, H. L. Bethem, K. S. E. Eikema, E. J. Salumbides, W. Ubachs, M. Beyer, N. J. Holsch, J. A. Agner, F. Merkt, L.-G Tao,S.-M Hu and Ch. Jungen, Phys. Rev. Lett.{\bf 121}, 013001(2018)
\bibitem(33)B. Jeziorski and W Kolos, Chem. Phys. Lett.,{\bf 3},678(1969)
\bibitem{34}Y. Hijikata,H. Nakashima and H. Nakatsuji, J. Chem Phys., {\bf 130},024102(2009)
\bibitem{35}M. Stanke, D.Kedziera, S. Bubin, M. Molski and L. Adamiwicz, J Chem Phys., {\bf 128}, 114313(2008)
\bibitem{36}S. Bubin, F. Leonarski, M Stanke and L Adamowicz, Chem. Phys. Lett.,{\bf 477},12(2009)
\bibitem{37}B. Chen and J. B. Anderson, J Chem Phys.,{\bf 102},2802(1995)
\bibitem{38}L. M. Wang and Z. C. Yan,Phys. Rev. A,{\bf 97}, 060501(2018)
\bibitem{39}R. Pino and V. Mujica, J. Phys. B,{\bf 31}4537(1998)
\bibitem{40}S. A. Alexander and R. L. Coldwell, J. Chem. Phys.,{\bf 129},114306(2008)
\bibitem{41}A. W. Jasper, C. Zhu, S. Nangia and D. G. Truhlar, Faraday Discuss, {\bf 127}, 1-22(2004)
\bibitem{42}M. D. Donsker and M. Kac, J. Res. Natl. Bur. Stand {\bf 44},
581 (1950); see also, M.Kac, in Proceedings of the Second Berkeley Symposium
(Berkeley Press, California (1951)).
\bibitem{43} A. Korzeniowski, J.L. Fry, D. E. Orr and N. G. Fazleev, Phys Lett {\bf 69}, 893,1992
\bibitem{44} R. P. Feynman, Rev. Mod. Phys.{\bf 20},367(1948)
\bibitem{45} R. P. Feynman and A. R. Hibbs, Quantum Mechanics and Path Integrals(McGraw-Hill,NY(1965))
\bibitem{46} H. S. Schulman, Techniques and Applications of Path Integrations,(Wiley, NY, 1993)
\bibitem{47} P. Exner, Open Quantum Systems and Feynman Integrals(Reidal Pub. Co., Boston, MA, 1985)
\bibitem{48} T Kato, Commun. Pure Appl Math,{\bf 10},151(1957)
\bibitem{49} M. D. Donsker and S. R. Varadhan, in Proc. of the International
Conference on Function space Integration ( Oxford Univ. Press 1975)pp. 15-33.
\bibitem{50} N. Wiener, J. Math and Phys., {\bf 2},132,(1923).
\bibitem{51} G. Roepstorff, Path integral approach to quantum Physics(Springer, 1994)
\bibitem{52} S. B. Doma, M. Abu-Shady, F. N. EI-Gammal and A. A. Amer, Molecular Physics{\bf 114},1787(2016)
\bibitem{53} Sumita Datta, S. A. Aleaxnder and R. L. Coldwell, Int. J. Q. Chem.,{\bf 112},731(2012)
\bibitem{54} J. S. Sims and S. A. Hagstrom, J. Chem Phys.,{\bf 124},094101(2006)
\bibitem{55} S.A. Alexander and R. L. Coldwell, J. Chem. Phys.,{\bf 121},11557(2004)
\bibitem{56} S. Datta, S. A. Alexander and R. L. Coldwell, Int. J. Q. Chem, {\bf 111}, 4106,(2011)
\bibitem(57) W. Kolos, J. Chem. Phys., {\bf 101}(1994)
\bibitem{58} N. M. Tubman, I.Kylanpaa, S. H. Hammes-Schiffer, D. M. Ceperley, Phys. Rev A 90, 042507(2014)
\bibitem{59} Y. P. Sarwono, F. U. Rahman and R Zhang, New J. Phys.,{\bf 22},093059(2020)
\bibitem{60} B. Simon, Functional Integrals and Quantum Mechanics(Academic Press, NY, 1979).
\bibitem{61} R. Griego and R. Hersh, Theory of random evolutions with applications to partial differential equations, Transact. Am. Math. Soc
{\bf 156}, 405(1971)
\bibitem{62}P. Billingsley,Convergence of Probability Measures; Wiley:New York, USA,1968
\end{thebibliography}
\end{document}